\documentclass[11pt,a4paper,english,superscriptaddress,prd,aps,preprint]{revtex4}
\usepackage{amssymb}
\usepackage{amsmath} 
\usepackage{slashed}
\usepackage{graphicx}
\usepackage{babel}
\usepackage{hyperref}
\usepackage{color}
\makeatletter
\AtBeginDocument{\let\@elt\relax}\makeatother

\begin{document}

\title{Two-loop renormalization of the CPT-even Lorentz-violating Scalar QED}

\author{L. C. T. Brito}
\email{lcbrito@dfi.ufla.br}
\affiliation{Departamento de F\'{i}sica, Instituto de Ci\^{e}ncias Naturais,  Universidade Federal de Lavras, Caixa Postal 3037,
	37200-900, Lavras, Minas Gerais, Brasil}

\author{J. C. C. Felipe}
\email{jean.cfelipe@ufvjm.edu.br}
\affiliation{Instituto de Engenharia, Ci\^{e}ncia e Tecnologia, Universidade Federal dos Vales do Jequitinhonha e Mucuri, Avenida Um, 4050 - 39447-790 - Cidade Universit\'{a}ria - Jana\'{u}ba, Minas Gerais, Brazil}

\author{A. C. Lehum}
\email{lehum@ufpa.br}
\affiliation{Faculdade de F\'{i}sica, Universidade Federal do Par\'{a}, 66075-110, Bel\'{e}m, Par\'a, Brazil}

\author{A. Yu. Petrov}
\email{petrov@fisica.ufpb.br}
\affiliation{Departamento de F\'{i}sica, Universidade Federal da Para\'{i}ba, Caixa Postal 5008,
	58051-970 Jo\~{a}o Pessoa, Para\'iba, Brazil}

\begin{abstract}
Investigating quantum effects arising from high loops in perturbation theory is crucial for the physical applications of any quantum field theory. This paper presents a comprehensive analysis of the two-loop renormalization of CPT-even Lorentz-violating scalar electrodynamics at the first order in the background vectors. We provide results for the self-energies of the photon and scalar field, as well as for the three-point function associated with the scalar-scalar-photon vertex, ensuring a thorough examination of the quantum effects. The calculations satisfy the ward identities, demonstrating their consistency. Computational tools were employed to carry out the calculations, and we provide additional details in the Supplemental Material for interested readers. Our contribution presents, for the first time, a two-loop calculation within the framework of the Lorentz-violating Standard Model Extension.
\end{abstract}

\maketitle

\section{Introduction}

Currently, the calculations of physical quantities in particle physics rely on the utilization of perturbative methods within the realm of quantum field theory. These methods enable precise calculations in high-order perturbation theory, which has been crucial for both testing the Standard Model (SM) and investigating its limitations and possible extensions \cite{Donoghue,Langacker:2017uah,Freitas}. 

An example that emphasizes the significance of high-order quantum corrections in particle physics is the anomalous magnetic moment \cite{Aoyama:2020ynm}. It is one of the observables used for precision tests of the Standard Model and has been calculated for the electron up to the five-loop order in QED \cite{Aoyama:2012wj}. Effects from the electroweak sector on the anomalous magnetic moment of leptons have been calculated up to the two-loop order \cite{Czarnecki:1995sz}, including the effects of strong interactions in the case of muons \cite{Czarnecki:2002nt}. Additionally, observables such as the anomalous magnetic moment are measured with high precision and may be sensitive to new physics \cite{Aoyama:2020ynm}. Therefore, this example suggests that any proposal of extending the SM should include precise analyses of high-order perturbation theory as part of its research program.  

In this paper we present, for the first time, a complete two-loop calculation in the realm of  Lorentz-violating  Standard Model Extension (LV SME).  Our purpose is to provide a comprehensive investigation into the renormalizability of a specific sector of the model, the CPT-even Lorentz-violating scalar quantum electrodynamics. The Lorentz-breaking modifications for various field theory models, formulated within the framework of LV SME, were  originally presented in \cite{ColKost1,ColKost2}. Within this theory, the main attention was paid to LV extensions of spinor QED and QCD, studied in hundreds of papers; see for a review, e.g. \cite{KosMew,KosMew1,ourrev} and references therein.

At the same time, LV extensions of scalar QED and scalar QCD, although they contribute to the Higgs sector of LV SME \cite{ColKost1,ColKost2}, are studied to a much lesser extent. Up to now, the main results regarding to perturbative treating of these theories are obtaining of the tree-level scattering amplitude in the LV extension of the Yukawa theory \cite{Altscal}, obtaining the lower one-loop contributions to gauge and scalar sector within the Abelian Higgs model \cite{Scarp2013}, and performing the one-loop renormalization of LV scalar QED \cite{BaetaScarpelli:2021dhz,Altschul:2022isc} and QCD \cite{LVQCD23}. Moreover, higher-loop corrections were not considered in scalar LV QED at all (and in the spinor one, they were treated only in \cite{noterm}, where absence of higher-loop contributions to the Carroll-Field-Jackiw (CFJ) term was proved). Certainly, study of higher-loop effects in scalar QED is a very interesting task. In this paper, we pursue this problem. Our starting point is the CPT-even LV QED involving LV terms both in scalar and gauge sectors, and within it, we calculate two-loop contributions to two- and three-point functions, so that, using the gauge symmetry requirement, we can easily recover the result for the four-point gauge-scalar function. It ensures that the renormalization of the model at two-loop level was thoroughly considered. 

The structure of the paper looks like follows. In the section 2, we write down our model. In the section 3, the results for the two-point function of the gauge field are presented. In the section 4, we present the results for the two-point function of the scalar field, and in the section 5 for the three-point gauge-scalar function. The results are discussed in the section 6. In addition to the content in the paper, the interested readers can find Supplemental Material at their disposal, which provides details about the calculation.

Throughout this paper, we use natural units $c=\hbar=1$ and $(+---)$ as the spacetime signature. 

\section{The massless CPT-even Lorentz-violating Scalar electrodynamics}\label{sec01}

Let us consider the model described by the bare Lagrangian~\cite{BaetaScarpelli:2021dhz}
\begin{eqnarray}\label{eq01}
\mathcal{L}&=&  (D_{\mu}\phi)^\dagger\left(\eta^{\mu\nu}+c^{\mu\nu}\right) D_{\nu}\phi - \frac{\lambda_0}{4} (\phi^\dagger\phi)^2 -\frac{1}{4}F^{\mu\nu}F_{\mu\nu}+\frac{1}{4}\kappa_{\mu\nu\alpha\beta}F^{\mu\nu}F^{\alpha\beta},
\end{eqnarray}
\noindent where $D_{\mu}=\partial_{\mu} - ieA_{\mu}$ is the covariant derivative, $c^{\mu\nu}$ and $\kappa_{\mu\nu\alpha\beta}$ are Lorentz-violating CPT-even dimensionless constant tensors.

Our focus in this study is on investigating the ultraviolet (UV) properties of the model. Therefore, we do not concern ourselves with the infrared (IR) divergences that may arise in the massless model. As usual, infrared problems can be avoided in the intermediary stages of the calculation by introducing a mass regulator in the field propagators.

Let us consider the assumption that $c^{\mu\nu}={Q_1}_0 u^\mu u^\nu$ represents an aether-like Lorentz-violating tensor, where $u^\mu$ is a  light-like four-vector with a magnitude of unity. We express $\kappa_{\mu\nu\alpha\beta}$ in terms of $c_{\mu\nu}$ as follows 
\begin{eqnarray}
\kappa_{\mu\nu\alpha\beta}&=&\frac{{Q_2}_0}{{Q_1}_0}\left(c_{\mu\alpha}\eta_{\nu\beta}-c_{\mu\beta}\eta_{\nu\alpha}+\eta_{\mu\alpha}c_{\nu\beta}-\eta_{\mu\beta}c_{\nu\alpha}\right),
\end{eqnarray}
\noindent doing with that our Lagrangian (\ref{eq01}) becomes
\begin{eqnarray}\label{eq03}
\mathcal{L}&=&  (D_{\mu}\phi)^\dagger\left(\eta^{\mu\nu}+{Q_1}_0u^{\mu}u^{\nu}\right) D_{\nu}\phi - \frac{\lambda_0}{4} (\phi^\dagger\phi)^2 -\frac{1}{4}F^{\mu\nu}F_{\mu\nu}+{Q_2}_0u_{\mu}u_{\nu}F^{\mu\alpha}{F^{\nu}}_{\alpha}.
\end{eqnarray}

In order to define the renormalization of the model let us write the counterterms. We start defining the quantum fields as $\phi\rightarrow Z_2^{1/2}\phi$ and $A^\mu\rightarrow Z_3^{1/2}A^\mu$, so
\begin{eqnarray}\label{eq04}
\mathcal{L}&=&  Z_2 (\partial^{\mu}\phi)^\dagger\partial_{\mu}\phi - \frac{\lambda Z_\lambda}{4} (\phi^\dagger\phi)^2 -\frac{Z_3}{4}F^{\mu\nu}F_{\mu\nu}+ie Z_1 A^\mu\left(\phi^\dagger\partial_\mu \phi-\phi\partial_\mu \phi^\dagger \right) \nonumber\\
&&+e^2Z_4\phi^\dagger\phi A^\mu A_\mu +u_{\mu}u_{\nu}\left[Q_1Z_5 (\partial_{\mu}\phi)^\dagger\partial_{\nu}\phi+ {Q_2} Z_6 F^{\mu\alpha}{F^{\nu}}_{\alpha}\right]\nonumber\\
&&+u^\mu u^\nu \left[iQ_1eZ_7 A_\nu \left(\phi^\dagger\partial_\mu \phi-\phi\partial_\mu \phi^\dagger \right) +Q_1e^2Z_8\phi^\dagger\phi A^\nu A_\mu \right] +\mathcal{L}_{GF}+\mathcal{L}_{CT}
\end{eqnarray}
\noindent where 
\begin{eqnarray}
Z_1e&=&\mu^{-2\epsilon}e_0Z_2Z_3^{1/2},\,\,\,\,\,\,e^2Z_4=\mu^{-2\epsilon}e_0^2Z_2Z_3,\,\,\,\,\,\ \lambda Z_\lambda=\mu^{-2\epsilon}\lambda_0 Z_2^2,\,\,\,\,\,\, Q_1Z_5=\mu^{-2\epsilon}{Q_1}_0Z_2, \nonumber\\
Q_2Z_6&=&\mu^{-2\epsilon}{Q_2}_0Z_3,\,\,\,\,\,\,Q_1eZ_7=\mu^{-2\epsilon}{Q_1}_0 e_0Z_2Z_3^{1/2},\,\,\,\,\,\, Q_1e^2Z_8=\mu^{-2\epsilon}{Q_1}_0e_0^2Z_2Z_3.
\label{def_counteterms}
\end{eqnarray}
Here, $\mu$ is the dimensional regularization mass scale, while $\epsilon=(4-D)/2$ with $D$ being the dimension of spacetime.

We will now discuss the one-loop corrections to the scalar and photon self-energies, and three-point function of the model as well, with the aim of computing the corresponding $Z$ factors. To achieve this, we adopt the approach of expanding any counterterm $Z_i$ as a power series of the coupling constants, determining them order by order within the perturbative expansion. Specifically, we express $Z_i$ as
\begin{eqnarray}
Z_i = 1 + Z^{(1)}_i + Z^{(2)}_i + \cdots,
\end{eqnarray}
where $Z_{i}^{(n)}$ represent de counterterm at $n$-loops.
In order to calculate all the relevant functions, we utilize a modified version of a suite of MATHEMATICA packages~\cite{feyncalc,feynarts,feynrules,Shtabovenko:2016whf}. We also utilize the Tarasov algorithm \cite{tarasov} to reduce two-loop integrals into a set of basic ones. The implementation of the Tarasov algorithm is achieved through the TARCER package \cite{tarcer}, where the basic integrals was computed in \cite{tsil}. These computational tools facilitate the computation and manipulation of Feynman diagrams and associated quantities in our analysis.

\section{Calculation of the two-loop photon self-energy} \label{sec2-twoloop}

\begin{figure}[ht]
	\includegraphics[angle=0,width=16cm]{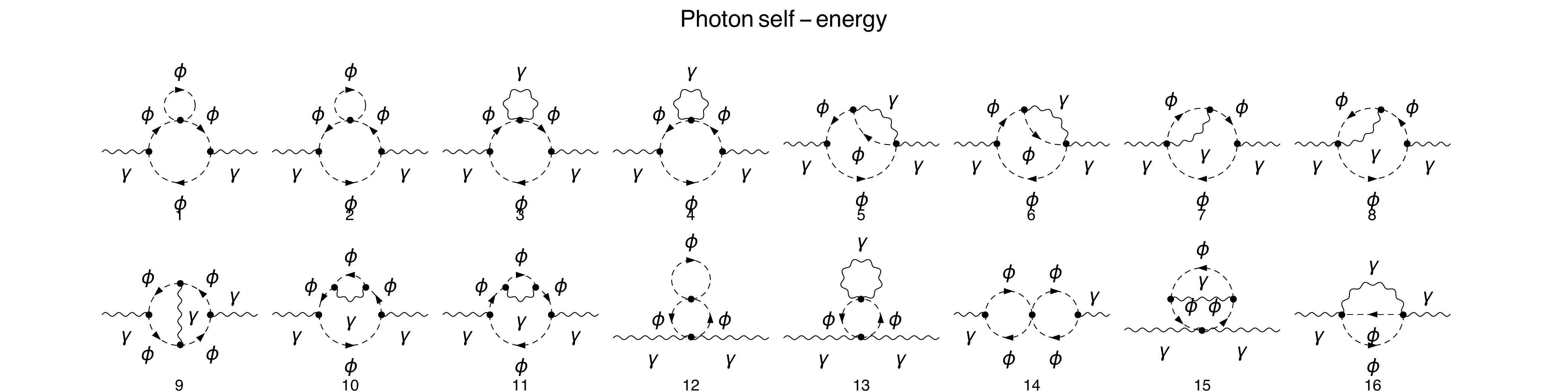}
	\caption{Feynman diagrams for the two-loop photon self-energy. Dashed and wavy lines represent the scalar and photon propagators, respectively.}
	\label{fig01}
\end{figure}

\begin{figure}[ht]
	\includegraphics[angle=0 ,width=12cm]{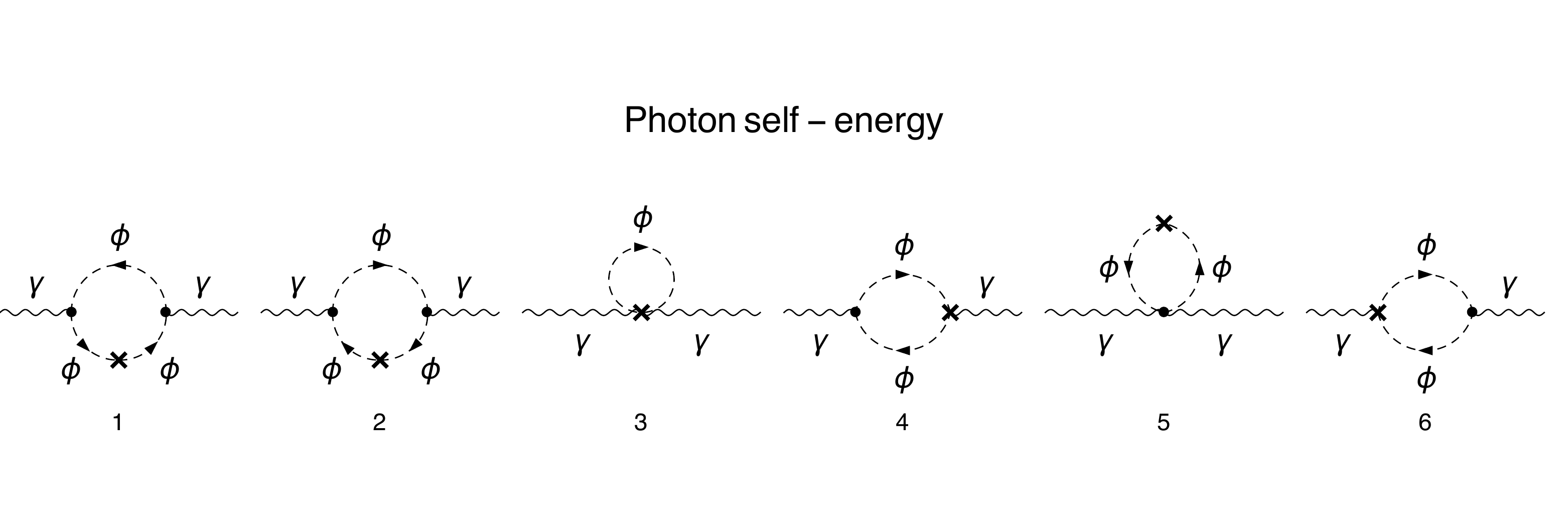}
	\caption{Contribution of order $e^4$ to the photon self-energy. The crossed vertices represent the insertion of an one-loop counterterm.}
	\label{fig02}
\end{figure}

\begin{figure}[ht]
	\includegraphics[angle=0 ,width=16cm]{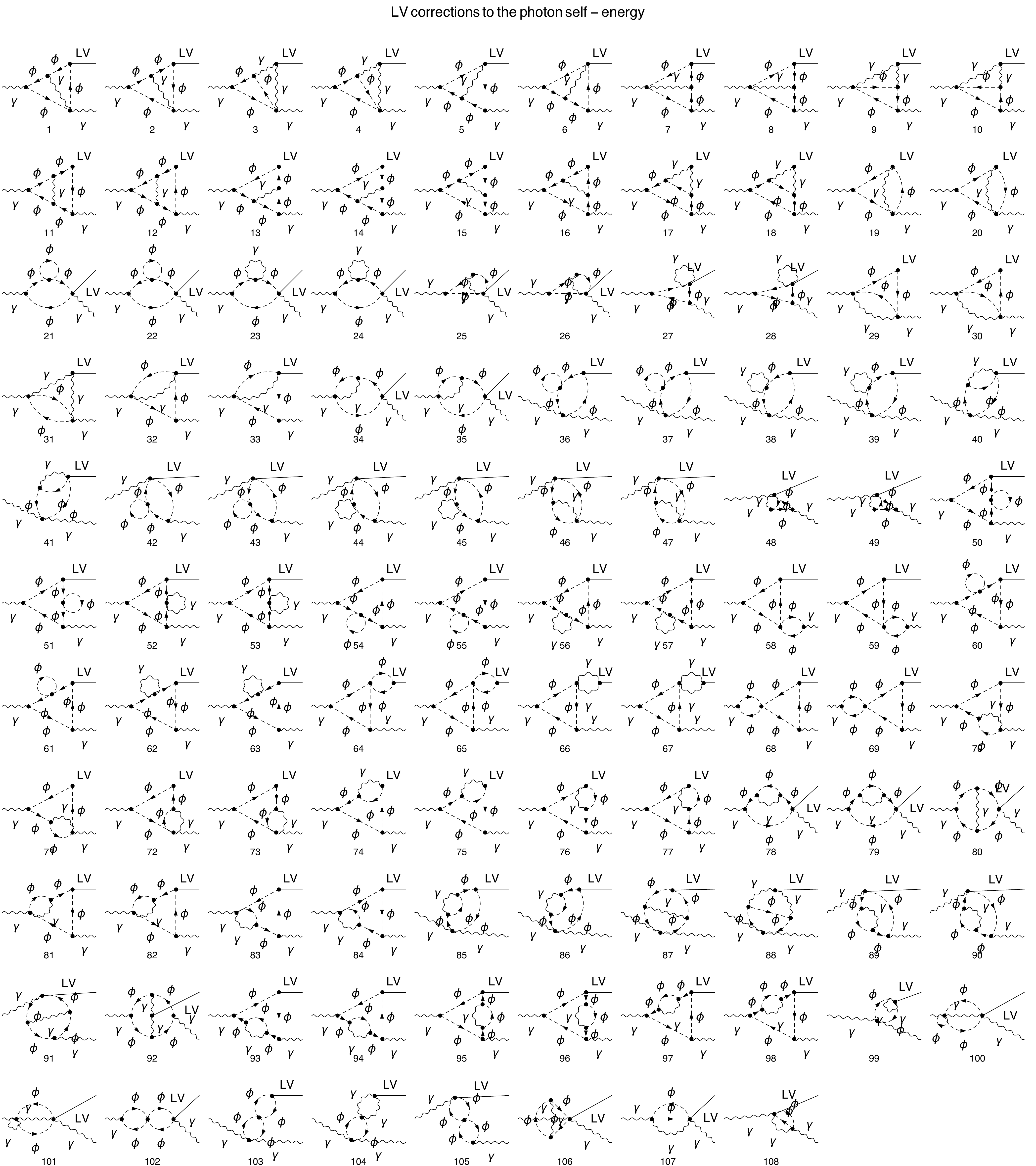}
	\caption{Feynman diagrams for the LV corrections to the two-loop photon self-energy. The straight line represents the insertion of a LV vertex.}
	\label{fig03}
\end{figure}

\begin{figure}[ht]
\includegraphics[angle=0 ,width=16cm]{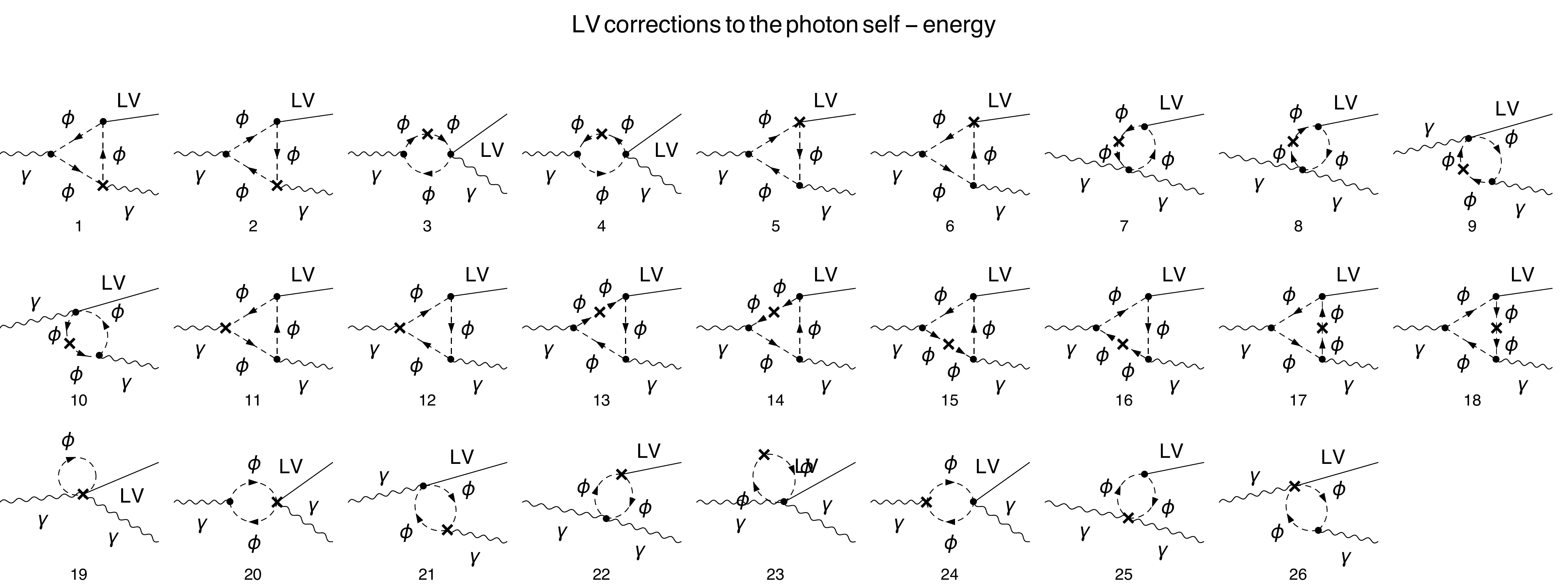}
	\caption{One-loop counterterm insertions in the LV corrections of order $e^4$ to the photon self-energy.}
	\label{fig04}
\end{figure}

We begin by examining the two-loop corrections to the photon self-energy. The two-loop polarization tensor can be separated into two components. The first component is the conventional term, proportional to $\left(p^2 \eta^{\mu \gamma}-p^\mu p^\gamma\right)$, depicted in Figs. \ref{fig01} and \ref{fig02}. The second component corresponds to the insertion of a Lorentz-violating (LV) vertex, as illustrated in Figs. \ref{fig03} and \ref{fig04}.

The first component has the following expression 
\begin{eqnarray}
\Pi^{\mu\gamma}_{2l}(p)&& =-i \left(p^2 \eta^{\mu  \gamma }-p^{\mu } p^{\gamma }\right) \Big[\frac{e^4}{128 \pi^4 \epsilon}+Z_3^{(2)}\nonumber\\
&&-\frac{e^2 (Z_1^{(1)}-Z_2^{(1)})}{72 \pi^2 \epsilon } \left(\epsilon  (3 \gamma -8-\log 64 -3 \log\pi )+3 \epsilon \log\left(-p^2\right) -3\right) \Big]+\mathrm{finite}.
\end{eqnarray}
\noindent Since $Z_1^{(1)}=Z_2^{(1)}$ (Ward identity), we have
\begin{eqnarray}
Z_3^{(2)}&& =-\frac{e^4}{128 \pi^4 \epsilon}.
\end{eqnarray}
\noindent The calculation of all the one-loop counterterms was previously carried out in Ref. \cite{Altschul:2022isc}, and we utilize those results to determine the amplitude depicted in Fig. \ref{fig02}.

The diagrams representing the Lorentz-violating (LV) contribution to the two-loop polarization tensor are shown in Figs. \ref{fig03} and \ref{fig04}. The computation of all the one-loop counterterms was previously performed in Ref. \cite{Altschul:2022isc}, and we employ those results to determine the amplitude depicted in Fig. \ref{fig04}. The explicit expression for this amplitude can be found in the Supplemental Material, where a more detailed calculation is provided. The corresponding expression is given by
\begin{eqnarray}
-i\Pi^{\mu\gamma}_{2l}(p)&& = (p\cdot u) \left[\eta^{\mu\gamma} (p\cdot u )-p^{\mu } u^{\gamma }+u^{\mu } \left(p^2 u^{\gamma }-p^{\gamma } (p\cdot u)\right)\right]\nonumber\\
&&\times\left( -\frac{e^4 (2 Q_1+Q_2)}{1536 \pi^4\epsilon^2}
-\frac{e^4 (38 Q_1+Q_2)}{4608 \pi^4 \epsilon}
+\frac{ Z_{6}^{(2)}Q_2}{2}
\right)+\mathrm{finite}.
\end{eqnarray}
\noindent  Imposing finiteness, we find
\begin{eqnarray}
Z_{6}^{(2)}=\frac{e^4 (2 Q_1+Q_2)}{768 \pi^4Q_2 \epsilon^2}
+\frac{e^4 (38 Q_1+Q_2)}{2304 \pi^4 Q_2 \epsilon}.
\end{eqnarray}
\noindent Notice the presence of a double pole in $Z_{6}^{(2)}$, characteristic of the two-loop correction.

Upon referring to the one-loop result in \cite{Altschul:2022isc}, we obtain the following expressions for the counterterms
\begin{subequations} \begin{eqnarray}\label{1-loop_Z3_sqed} Z_3 &=& 1+Z_3^{(1)}+Z_3^{(2)}+\cdots= 1-\frac{e^2}{48 \pi^2 \epsilon}-\frac{e^4}{128 \pi^4 \epsilon}+\cdots,\\ 
Z_6 &=& 1+Z_6^{(1)}+Z_6^{(2)}+\cdots = 1+\frac{{e}^2 {Q_1}}{24 \pi^2 Q_2 \epsilon}+\frac{e^4 (2 Q_1+Q_2)}{768 \pi^4Q_2 \epsilon^2}
+\frac{e^4 (38 Q_1+Q_2)}{2304 \pi^4 Q_2 \epsilon}+\cdots. \end{eqnarray} \end{subequations}

\section{Calculation of the two-loop scalar field self-energy} \label{sec3-twoloop} 

\begin{figure}[ht]
\includegraphics[angle=0 ,width=16cm]{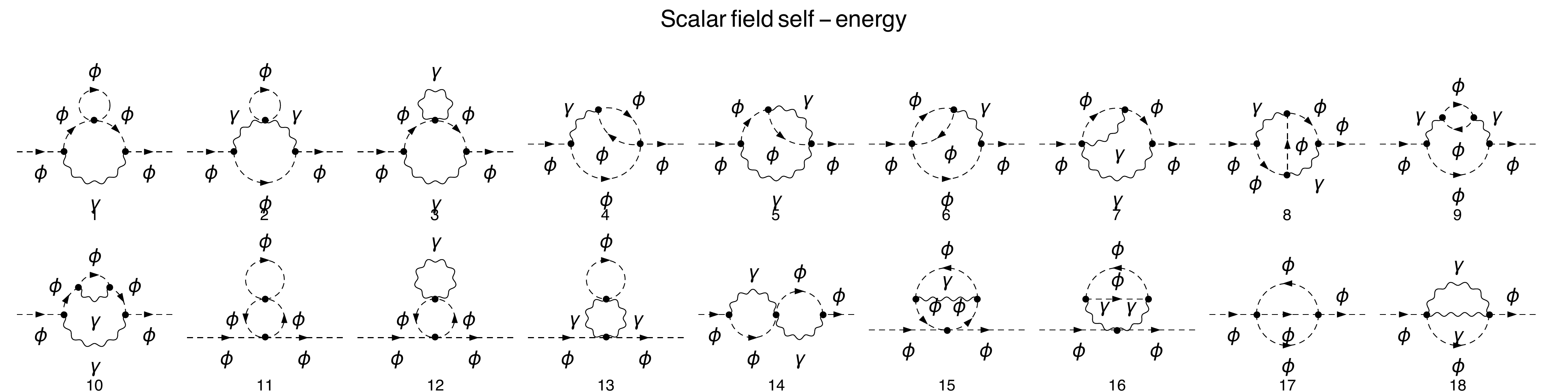}
	\caption{Scalar field self-energy.}
	\label{fig05}
\end{figure}

\begin{figure}[ht]
\includegraphics[angle=0 ,width=16cm]{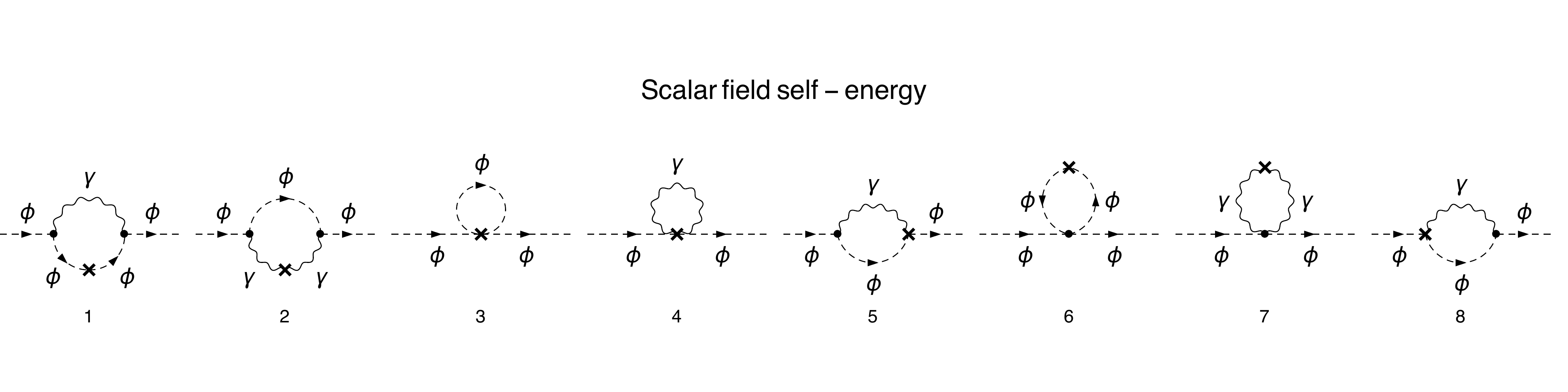}
	\caption{Scalar field self-energy.}
	\label{fig06}
\end{figure}

\begin{figure}[ht]
\includegraphics[angle=0 ,width=16cm]{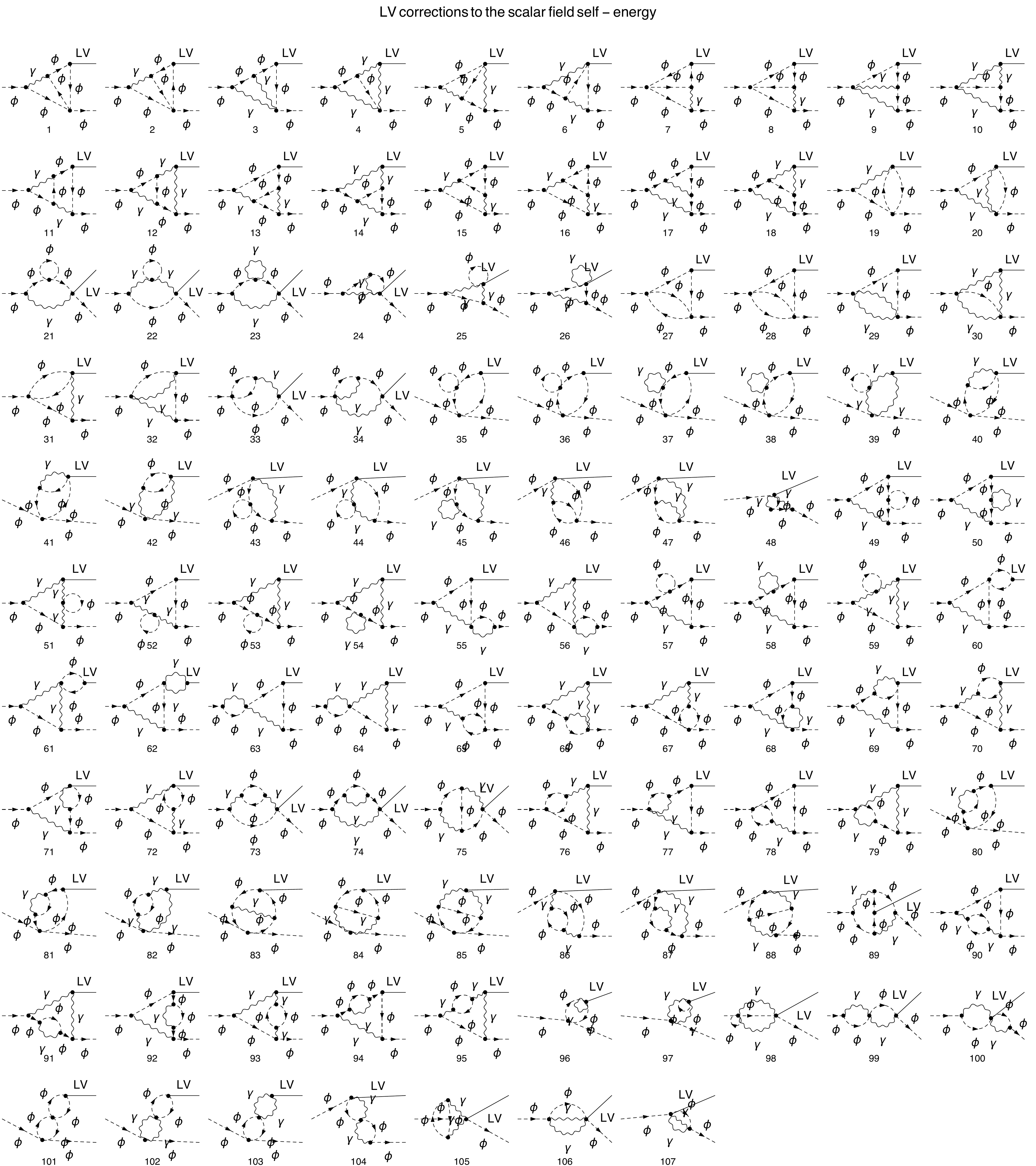}
	\caption{LV corrections to the scalar field self-energy.}
	\label{fig07}
\end{figure}

\begin{figure}[ht]
\includegraphics[angle=0 ,width=16cm]{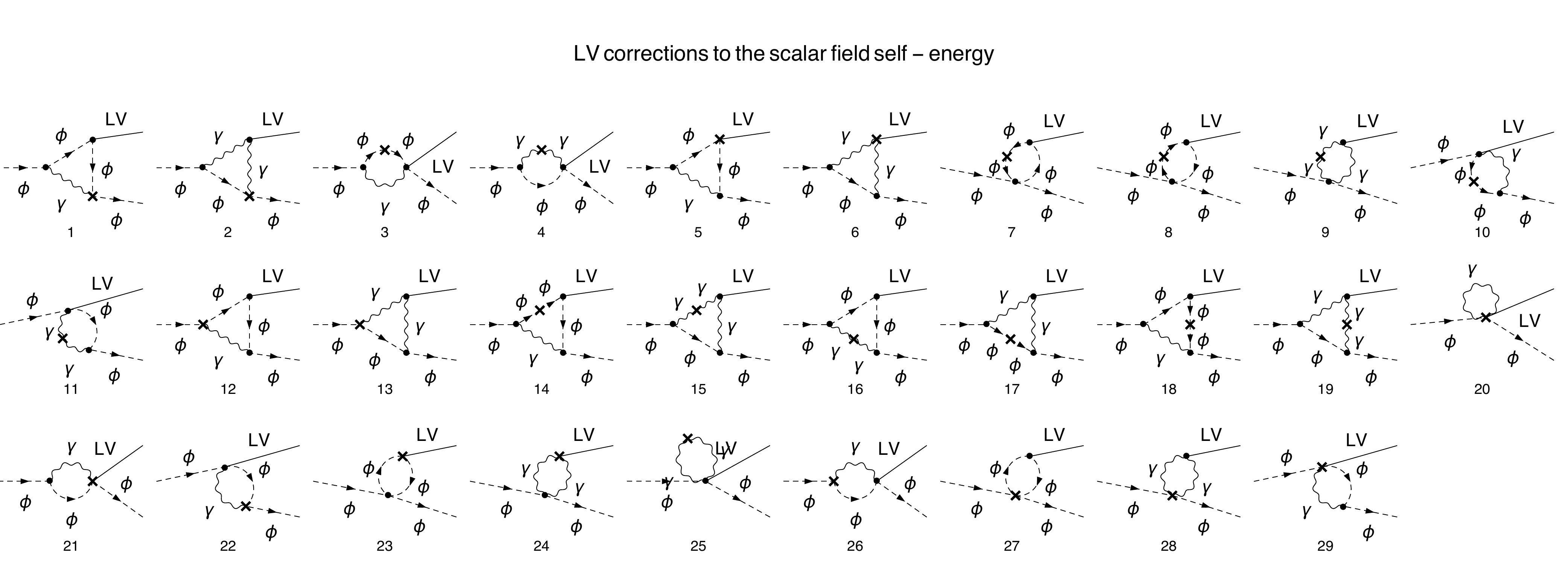}
	\caption{One-loop counterterm insertions in the LV corrections to the scalar field self-energy.}
	\label{fig08}
\end{figure}

Let us proceed with the computation of the two-loop scalar self-energy depicted in Figs. \ref{fig05} to \ref{fig08}. The expression for the conventional amplitude, as shown in Figs. \ref{fig05} and \ref{fig06}, is given by
\begin{eqnarray}\label{eqsse01}  
-i\Sigma_2(p) &=&p^2\left( -\frac{5e^4}{512 \pi^4\epsilon^2} + \frac{ 40 e^4+3 \lambda ^2}{6144 \pi^4\epsilon} + Z_2^{(2)} \right) +\text{finite}. 
\end{eqnarray}

The diagrams depicting the insertion of the Lorentz-violating (LV) parameters at first order are shown in Figs. \ref{fig07} and \ref{fig08}. The corresponding expression for this contribution is given by:
\begin{eqnarray}\label{eqsse02}
 -i\Sigma_{LV}(p) &=&\left( -\frac{5 e^4 (11 Q_1+4 Q_2)}{1536 \pi^4 \epsilon^2} 
 +\frac{ 8e^4(13 e^4 Q_1- Q_2)+9 \lambda ^2 Q_1}{18432 \pi ^4 \epsilon }+Z_5^{(2)}Q_1\right) (p\cdot u)^2,
\end{eqnarray}
where finite contributions were omitted.

By imposing finiteness and including the one-loop contributions to the counterterms calculated in \cite{Altschul:2022isc}, we obtain the following expressions
\begin{subequations}
 \begin{eqnarray}
  Z_2 &=& 1+ Z_2^{(1)}+Z_2^{(2)}+\cdots=1+\frac{e^2 }{8 \pi^2 \epsilon}+\frac{5e^4}{512 \pi^4\epsilon^2} - \frac{ 40 e^4+3 \lambda ^2}{6144 \pi^4\epsilon}\cdots; \label{ctphi01}\\
  Z_5 &=& 1+Z_5^{(1)}+Z_5^{(2)}+\cdots\nonumber\\
  & =&  1+\frac{e^2 (4 Q_1+Q_2)}{16 \pi^2  Q_1\epsilon}+\frac{5 e^4 (11 Q_1+4 Q_2)}{1536 \pi^4 Q_1\epsilon^2}
  -\frac{ 8e^4(13 e^4 Q_1- Q_2)+9 \lambda^2 Q_1}{18432 \pi^4 Q_1\epsilon}+\cdots.\label{ctphi02}
 \end{eqnarray}
\end{subequations}

\noindent Notice the existence of a double pole in the counterterm factors $Z_2^{(2)}$ and $Z_5^{(2)}$.

\section{Calculation of the two-loop three-point function} \label{sec4-twoloop} 

Let us proceed with the computation of the three-point function. The expression for the conventional amplitude is given by
\begin{eqnarray}\label{eqsse01}  
-i\Gamma^{\mu}(p_1,p_2) &=& e(p_1^\mu-p_2^\mu)\left( -\frac{5e^4}{512 \pi^4\epsilon^2} + \frac{ 40 e^4+3 \lambda ^2}{6144 \pi^4\epsilon} + Z_1^{(2)} \right) +\text{finite}. 
\end{eqnarray}

The expression for the diagrams with the insertion of the LV parameters at first order is given by:
\begin{eqnarray}\label{eqsse02}
 -i\Sigma_{LV}(p) &=&e\left( -\frac{5 e^4 (11 Q_1+4 Q_2)}{1536 \pi^4 \epsilon^2} 
 +\frac{ 8e^4(13 e^4 Q_1- Q_2)+9 \lambda ^2 Q_1}{18432 \pi ^4 \epsilon }+Z_7^{(2)}Q_1\right) u^\mu \left((p_1-p_2)\cdot u\right)\nonumber\\
 &&  +\text{finite}.
\end{eqnarray}

By imposing finiteness and including the one-loop contributions to the counterterms calculated in \cite{Altschul:2022isc}, we obtain the following expressions
\begin{subequations}
 \begin{eqnarray}
  Z_1 &=& 1+Z_1^{(1)}+Z_1^{(2)}+\cdots=1+\frac{e^2 }{8 \pi^2 \epsilon}+\frac{5e^4}{512 \pi^4\epsilon^2} - \frac{ 40 e^4+3 \lambda ^2}{6144 \pi^4\epsilon}\cdots=Z_2; \label{cttpf01}\\
  Z_7 &=&1+Z_7^{(1)}+Z_7^{(2)}+\cdots\nonumber\\
  & =&  1+\frac{e^2 (4 Q_1+Q_2)}{16 \pi^2  Q_1\epsilon}+\frac{5 e^4 (11 Q_1+4 Q_2)}{1536 \pi^4 Q_1\epsilon^2}
  -\frac{ 8e^4(13 e^4 Q_1- Q_2)+9 \lambda^2 Q_1}{18432 \pi^4 Q_1\epsilon}+\cdots=Z_5.\label{cttpf02}
 \end{eqnarray}
\end{subequations}

\noindent Notice that we showed that $Z_1=Z_2$ and $Z_5=Z_7$ up to two-loop order,  which confirms the gauge covariance of our quantum corrections. From the relations (\ref{def_counteterms}), we can see that the counterterms $Z_4$ and $Z_8$ are determined by the counterterms $Z_2$ and $Z_3$. Therefore, the renormalization at the two-loop order is complete.

\section{Final Remarks}\label{summary}

 Let us discuss our results. Perturbative calculations involving high-loop Feynman diagrams are challenging in quantum field theory. This kind of calculation invariably requires powerful computational tools and is crucial for comparing theoretical predictions with experimental results in high-precision tests.  In this paper, we provide a two-loop analysis of the renormalization of scalar LV QED.  We calculated two-loop contributions to two- and three-point functions, both Lorentz invariant and Lorentz-breaking ones. The calculation was performed at the lowest order in the Lorentz-violation parameters. Actually, using the gauge covariance requirement, we can recover the values of the four-point scalar-vector function as well, therefore, we can conclude that we succeeded to perform the complete two-loop renormalization of the scalar LV QED. The main value of our result consists in the fact that it is actually the first ever performed explicit calculation of a two-loop correction in a LV field theory model.
  Also, this result in principle can be generalized to other LV theories such as CPT-odd LV scalar QED and LV scalar QCD, so, actually, our study opens new perspectives fir studies of LV field theory models. In particular, it is interesting to perform two-loop studies in spinor QED extending the results of \cite{KosPic} to the two-loop order. We expect to perform this study in a forthcoming paper.

\acknowledgments

The authors are grateful to J. R. Nascimento for important discussions.
The work of A. Yu.\ P. has been partially supported by the CNPq project No. 301562/2019-9.

\newpage

\begin{thebibliography}{99}

\bibitem{Donoghue}
John F. Donoghue, Eugene Golowich, and Barry R. Holstein, "Dynamics of the standard model", Cambridge Univ. Press, 2014.

\bibitem{Langacker:2017uah}
P.~Langacker,
``The Standard Model and Beyond,''
Taylor \& Francis, 2017.

\bibitem{Freitas}
Ayres Freitas, "Precision tests of the standard model", Theoretical Advanced Study Institute 2020" The Obscure Universe: Neutrinos and Other Dark Matters"-TASI2020; 1-26 June (2021): 5 [arXiv: 2012.11642 [hep-ph]].

\bibitem{Aoyama:2020ynm}
T.~Aoyama, N.~Asmussen, M.~Benayoun, J.~Bijnens, T.~Blum, M.~Bruno, I.~Caprini, C.~M.~Carloni Calame, M.~C\`e and G.~Colangelo, \textit{et al.}
``The anomalous magnetic moment of the muon in the Standard Model,''
Phys. Rept. \textbf{887} (2020), 1-166 [arXiv: 2006.04822 [hep-ph]].

\bibitem{Aoyama:2012wj}
T.~Aoyama, M.~Hayakawa, T.~Kinoshita and M.~Nio,
``Tenth-Order QED Contribution to the Electron g-2 and an Improved Value of the Fine Structure Constant,''
Phys. Rev. Lett. \textbf{109} (2012), 111807 [arXiv: 1205.5368 [hep-ph]].

\bibitem{Czarnecki:1995sz}
A.~Czarnecki, B.~Krause and W.~J.~Marciano,
``Electroweak corrections to the muon anomalous magnetic moment,'' Phys. Rev. Lett. \textbf{76} (1996), 3267-3270 [arXiv: hep-ph/9512369 [hep-ph]].

\bibitem{Czarnecki:2002nt}
A.~Czarnecki, W.~J.~Marciano and A.~Vainshtein,
``Refinements in electroweak contributions to the muon anomalous magnetic moment,''
Phys. Rev. D \textbf{67} (2003), 073006
[erratum: Phys. Rev. D \textbf{73} (2006), 119901].

\bibitem{ColKost1} D.~Colladay and V.~A.~Kostelecky,
``CPT violation and the standard model,''
Phys. Rev. D \textbf{55} (1997), 6760-6774
[arXiv:hep-ph/9703464 [hep-ph]].

\bibitem{ColKost2} D.~Colladay and V.~A.~Kostelecky,
``Lorentz violating extension of the standard model,''
Phys. Rev. D \textbf{58} (1998), 116002
[arXiv:hep-ph/9809521 [hep-ph]].

\bibitem{KosMew} V.~A.~Kostelecky and M.~Mewes,
``Signals for Lorentz violation in electrodynamics,''
Phys. Rev. D \textbf{66} (2002), 056005
[arXiv:hep-ph/0205211 [hep-ph]].

\bibitem{KosMew1} V.~A.~Kostelecky and M.~Mewes,
``Electrodynamics with Lorentz-violating operators of arbitrary dimension,''
Phys. Rev. D \textbf{80} (2009), 015020
[arXiv:0905.0031 [hep-ph]].

\bibitem{ourrev} A.~F.~Ferrari, J.~R.~Nascimento and A.~Y.~Petrov,
``Radiative corrections and Lorentz violation,''
Eur. Phys. J. C \textbf{80} (2020) no.5, 459
[arXiv:1812.01702 [hep-th]].

\bibitem{Altscal} B.~Altschul,
``Lorentz and CPT Violation in Scalar-Mediated Potentials,''
Phys. Rev. D \textbf{87} (2013) no.4, 045012
[arXiv:1211.6614 [hep-th]].

\bibitem{Scarp2013} L.~C.~T.~Brito, H.~G.~Fargnoli and A.~P.~Ba\^eta Scarpelli,
``Aspects of Quantum Corrections in a Lorentz-violating Extension of the Abelian Higgs Model,''
Phys. Rev. D \textbf{87} (2013) no.12, 125023
[arXiv:1304.6016 [hep-th]].

\bibitem{BaetaScarpelli:2021dhz}
A.~P.~Ba\^eta Scarpelli, J.~C.~C.~Felipe, L.~C.~T.~Brito and A.~Yu.~Petrov,
``One-loop calculations in CPT-even Lorentz-breaking scalar QED,''
Mod. Phys. Lett. A \textbf{37}, no.16, 2250100 (2022)
[arXiv:2111.14257 [hep-th]].

\bibitem{Altschul:2022isc}
B.~Altschul, L.~C.~T.~Brito, J.~C.~C.~Felipe, S.~Karki, A.~C.~Lehum and A.~Y.~Petrov,
``Three- and four-point functions in CPT-even Lorentz-violating scalar QED,''
Phys. Rev. D \textbf{107}, no.4, 045005 (2023)
[arXiv:2211.11399 [hep-th]].

\bibitem{LVQCD23} B.~Altschul, L.~C.~T.~Brito, J.~C.~C.~Felipe, S.~Karki, A.~C.~Lehum and A.~Y.~Petrov,
``Perturbative Aspects of CPT-Even Lorentz-Violating Scalar Chromodynamics,''
Phys. Rev. D \textbf{107}, no.11, 115002 (2023)
[arXiv:2304.03025 [hep-th]].

\bibitem{noterm} L.~C.~T.~Brito, J.~C.~C.~Felipe, A.~Y.~Petrov and A.~P.~Ba\^eta Scarpelli,
``No radiative corrections to the Carroll\textendash{}Field\textendash{}Jackiw term beyond one-loop order,''
Int. J. Mod. Phys. A \textbf{36} (2021) no.05, 2150033
[arXiv:2005.04637 [hep-th]].

\bibitem{feyncalc} 
R.~Mertig, M.~Bohm and A.~Denner,
``FEYN CALC: Computer algebraic calculation of Feynman amplitudes,''
Comput. Phys. Commun. \textbf{64}, 345-359 (1991)
V.~Shtabovenko, R.~Mertig and F.~Orellana,
``New Developments in FeynCalc 9.0,''
Comput. Phys. Commun. \textbf{207}, 432-444 (2016)
[arXiv:1601.01167 [hep-ph]];
V.~Shtabovenko, R.~Mertig and F.~Orellana,
``FeynCalc 9.3: New features and improvements,''
Comput. Phys. Commun. \textbf{256}, 107478 (2020).
[arXiv:2001.04407 [hep-ph]];

\bibitem{feynarts}
T.~Hahn,
``Generating Feynman diagrams and amplitudes with FeynArts 3,''
Comput. Phys. Commun. \textbf{140}, 418-431 (2001)
[arXiv:hep-ph/0012260 [hep-ph]].
  
\bibitem{feynrules}
A.~Alloul, N.~D.~Christensen, C.~Degrande, C.~Duhr, and B.~Fuks,
``FeynRules  2.0 - A complete toolbox for tree-level phenomenology,''
Comput. Phys. Commun. \textbf{185}, 2250 (2014).

\bibitem{Shtabovenko:2016whf}
V.~Shtabovenko,
``FeynHelpers: Connecting FeynCalc to FIRE and Package-X,''
Comput. Phys. Commun. \textbf{218}, 48-65 (2017)
[arXiv:1611.06793 [physics.comp-ph]].

\bibitem{tarasov}
O.~V.~Tarasov,
``Generalized recurrence relations for two loop propagator integrals with arbitrary masses,''
Nucl. Phys. B \textbf{502}, 455-482 (1997)
[arXiv:hep-ph/9703319 [hep-ph]].

\bibitem{tarcer}
R.~Mertig and R.~Scharf,
``TARCER: A Mathematica program for the reduction of two loop propagator integrals,''
Comput. Phys. Commun. \textbf{111}, 265-273 (1998)
[arXiv:hep-ph/9801383 [hep-ph]].

\bibitem{tsil}
S.~P.~Martin and D.~G.~Robertson,
``TSIL: A Program for the calculation of two-loop self-energy integrals,''
Comput. Phys. Commun. \textbf{174}, 133-151 (2006)
[arXiv:hep-ph/0501132 [hep-ph]].

\bibitem{KosPic} V.~A.~Kostelecky, C.~D.~Lane and A.~G.~M.~Pickering,
``One loop renormalization of Lorentz violating electrodynamics,''
Phys. Rev. D \textbf{65} (2002), 056006
[arXiv:hep-th/0111123 [hep-th]].


\end{thebibliography}
\end{document}